\documentclass[twocolumn,showpacs,preprintnumbers,amsmath,amssymb]{revtex4}
\usepackage{graphicx}
\usepackage{dcolumn}
\usepackage{bm}
\begin{document}


\title{Tunnelling of condensate magnetization in a double-well potential}

\author{\"O. E. M\"ustecapl{\i}o\u{g}lu,$^1$ M. Zhang,$^2$ and L. You$^{3}$}
\address{$^1$Ko\c{c} University, Department of Physics,
Rumelifeneri Yolu, 34450 Sar{\i}yer, Istanbul, Turkey}
\address{$^2$Center for Advanced Study, Tsinghua University,
Beijing, 100084, China}
\address{$^3$School of Physics, Georgia Institute of Technology, Atlanta GA
30332, USA}
\date{\today}

\begin{abstract}
We study quantum dynamical properties of a spin-$1$ atomic
Bose-Einstein condensate in a double-well potential. Adopting a mean
field theory and single spatial mode approximation, we characterize
our model
system as two coupled
 spins. For certain initial states, we find
full magnetization oscillations between wells not accompanied by
mass (or atom numbers) exchange. We identify
dynamic regimes of collective spin variables
arising from nonlinear self-interactions that are
different from the usual Josephson oscillations.
We also discuss magnetization beats and incomplete oscillations of
collective spin variables other than the magnetization.
Our study points to an alternative
approach to observe coherent tunnelling of a condensate through a (spatial)
potential barrier.
\end{abstract}
  \maketitle
\narrowtext


In different branches of physics, various systems larger than atomic size
have been examined to have signatures of purely quantum effects
particularly in quantum coherence and
tunnelling \cite{leggett87,leggett85,garg93}. The macroscopic
quantum tunnelling of magnetization
(MQTM) \cite{chudnovsky98,gunther95,villain94} is such an effect
that has been vigorously sought after and has been
claimed to be observed in molecular nanomagnets of ${\rm
Mn}_{12}-{\rm acetate}$ and ${\rm Fe}_{8}$
compounds \cite{lis80,wieghart84}. The surprising event of MQTM was also
believed to occur in iron storage protein
ferritin \cite{awschalom92}. Apart from its fundamental
significance at the interface of quantum and classical realms,
MQTM in a large spin system is expected to play a key role in
realizing technological advances for future's quantum computers and for
the newly emerging field of spintronics. Being macroscopic
quantum objects themselves, optically trapped atomic Bose-Einstein
condensates (BECs) \cite{kurn,mike} are also proposed to be
promising candidates for MQTM \cite{pu02}, though strictly speaking,
one may classify such systems as mesoscopic as well.

In optical traps hyperfine spin degrees of freedom become
accessible and for a condensate of atoms in the $F=1$ hyperfine
ground state, its order parameter is a three component
($M_F=-1,0,+1$) spinor. The tunnelling associated with the
internal (spin) degree of freedom can be induced by a transverse
magnetic field when it exceeds a critical value depending on the
strength of magnetic spin dipole-dipole interaction \cite{pu02}.
Such Josephson oscillations with internal degrees of freedom can
be compared to the well-known superfluid $^3$He Josephson
oscillations observed in recent years \cite{davis}. The tunnelling
associated with the external (spatial) center of mass degree of
freedom has been demonstrated with cold atoms
\cite{niu,ens,haycock00} as well as atomic condensates in an
optical lattice \cite{mark}. The underlying strong correlations in
spin and motional degrees of freedom in such systems may
eventually lead to novel applications in spintronics. Beyond the
semi-classical macroscopic Josephson oscillation physics, two
tunnelling coupled condensates can also display many particle
quantum correlations when the strength of tunnelling coupling is
tuned relative to the typical mean field interaction strength
inside each well \cite{you}.  While this phenomenon is well
understood for scalar condensates, it is so far unexplored for
spin-1 condensates \cite{milburn,sipe,ho}. In this article, we
propose a different and more direct mechanism for observing
coherent condensate Josephson dynamics based on magnetization
oscillations due to an external (spatial) Josephson coupling of a
spin-1 condensate in a double well potential, building upon
previous studies of a spin 1/2 condensate in a double well
potential \cite{ashab01,law,leggett01}.

Our model system consists of a spin-1 condensate in a double-well
potential. When tunnelling through the
potential barrier is weak, the wave function effectively behaves
as a superposition of those localized in the left (L) and right (R) wells.
In the mean field approximation,
we write the three component order parameter as
$\psi_i(\vec{r},t)=\phi_{Li}(\vec{r})
\xi_i(t)+\phi_{Ri}(\vec{r})\eta_i(t)$ with
$i=-,0,+$ respectively for each Zeeman state $|M_F\rangle$.
$\phi_{\nu i}$ is the ground state spatial wave
function of the $i$-th spin component in the $\nu$-th well ($\nu$=L,R).
Validity of the mean field approximation is discussed in the conclusion part
of the paper.
As has been studied in great detail \cite{law98,ho98},
both spin-1 condensates of $^{87}$Rb and $^{23}$Na
atoms are dominated by spin symmetric interaction. This leads to the
wide use of single spatial mode approximation (SMA) \cite{law98}, whereby the
mode function is taken to be the same of all three components,
and is itself determined only from the symmetric interactions.
We will adopt this approximation as recent studies have fully
delineated its validity regime \cite{pu,law98,yi}.
Denote atomic density as $n_{\nu}(\vec{r})$, hence
we take $\phi_{\nu i}=\sqrt{n_\nu(\vec{r})}$, which leads to
\begin{eqnarray}
\Psi(\vec{r},t)=\sqrt{n_L(\vec{r})}\,\vec{\xi}(t)+
\sqrt{n_R(\vec{r})}\,\vec{\eta}(t),
\end{eqnarray}
where $\vec f_\nu$ ($\vec f_L=\vec{\xi}$ and $\vec f_R=\vec{\eta}$)
is the spin-1 spinor
associated with the condensate in the $\nu$-th well, and
should be interpreted as a column vector, while
$\vec f_\nu^\dag$ a row vector in our notation.

We follow the convention of Ref. \cite{law98} to split the total
Hamiltonian for an interacting spin-1 condensate into two parts
$H=H_S+H_A$. After integrating over the spatial variables and
adopting the SMA introduced above we get
\begin{eqnarray}
H_S&=&\sum_{\nu=L,R}\left(\epsilon_\nu\vec{f}_\nu^\dag\vec{f}_\nu
+\frac{1}{2}\lambda_\nu^{(S)}\vec{f}_\nu^\dag\vec{f}_\nu\,\vec{f}_\nu^\dag\vec{f}_\nu\right)
\nonumber\\
&&+J(\vec{\xi}^\dag\vec{\eta}+\vec{\eta}^\dag\vec{\xi}),\\
H_A&=&\frac{1}{2}\sum_{\nu=L,R}
{\lambda_\nu^{(A)}}\sum_{j=x,y,z}\vec{f}_\nu^\dag
F_j\,\vec{f}_\nu\,\vec{f}_\nu^\dag F_j\,\vec{f}_\nu,
\end{eqnarray}
where $F_j$ are spin-1 matrices. We have defined $\lambda_\nu^{(S/A)}=c_{S/A}\int
d\vec{r}\,n_{\nu}^2(\vec{r})$, the single well ground state energy
\begin{eqnarray}
\epsilon_{\nu}&=& \int d\vec{r}\left(\frac{\hbar^2}{2M}\left[\nabla\sqrt{n_\nu}\right]^2
+\sqrt{n_\nu}\,V\sqrt{n_\nu}\right),
\end{eqnarray}
and a positive tunnelling coefficient
\begin{eqnarray}
J&=& \int
d\vec{r}\left(\frac{\hbar^2}{2M}\nabla\sqrt{n_L}\cdot\nabla\sqrt{n_R}+\sqrt{n_L}\,V\sqrt{n_R}\right).
\quad
\end{eqnarray}
The spin symmetric and asymmetric coefficients are
$c_S={4\pi\hbar^2}(a_0+2a_2)/3M$ and
$c_A={4\pi\hbar^2}(a_2-a_0)/3M$ in terms of the scattering length
$a_0$ ($a_2$) in the total spin channel of 0 (2) of the two colliding spin-1 atoms.
The tunnelling coupling is the same for all spin components as the
potential is assumed spin independent, i.e.,
$V_{ij}=V\delta_{ij}$ for simplicity. The terms involving small
spatial overlaps between the wave functions in each well
are neglected \cite{milburn}.

The time dependent Schrodinger equations for the spinors are
obtained from $i\hbar\,{d\vec{\xi}}/{dt}={\delta
H(\vec{\xi},\vec{\eta},\vec{\xi}^\dag,\vec{\eta}^\dag)}/{\delta\vec{\xi}^\dag}$,
which gives in a compact form,
\begin{eqnarray}
i\hbar\frac{d \vec{\xi}}{dt} &=&
\left(\epsilon_L+(\lambda_L^{(S)}+\lambda_L^{(A)})|\vec{\xi}|^2-\lambda_L^{(A)}h_L\right)\,\vec{\xi}+J\,\vec{\eta},\nonumber \\
i\hbar\frac{d\vec{\eta}}{dt} &=&
J\,\vec{\xi}+\left(\epsilon_R+(\lambda_R^{(S)}+\lambda_R^{(A)})|\vec{\eta}|^2-\lambda_R^{(A)}h_R\right)\,
\vec{\eta}, \hskip 24pt \label{matrixform}
\end{eqnarray}
the equivalent coupled Gross-Pitaevskii equation for the
two spinors. We have defined a short hand notation
$h_\nu\equiv\vec{f'}_\nu^*\otimes\vec{f'}_\nu^T$
with $\vec{f'}^T=(f_-, -f_0, f_+)$.
Both $\vec{\xi}$ and $\vec{\eta}$ are normalized to unity, so we attempt to
find the stationary states satisfying
\begin{eqnarray}
i\hbar\frac{d}{dt}\left(\begin{array}{c}
            \vec{\xi} \\
            \vec{\eta}
            \end{array}\right)=\mu\left(\begin{array}{c}
                                                \vec{\xi} \\
                                                \vec{\eta}
                                                 \end{array}\right).
\end{eqnarray}

In this short article, we will focus on the tunnelling
dynamics in a symmetric double well with
$\epsilon_L=\epsilon_R=\epsilon,\lambda_L^{(S)}=\lambda_R^{(S)}=\lambda_S,
\lambda_L^{(A)}=\lambda_R^{(A)}=\lambda_A$. In this case,
inter-well tunnelling couples the same spin
components, while intra-well interactions cause spin
mixing, coupling components~$|+\rangle$ and $|-\rangle$ to
$|0\rangle$.
We have performed extensive simulations and
discovered a variety of interesting solutions:
self-trapping for each spinor component,
for combinations of different spinor components,
as well as general nonlinear Josephson type
complete oscillations. To illustrate these results, we will
selectively display some oscillation dynamics that
might be useful to guide ongoing experimental efforts.
For these solutions to be physically meaningful,
we have also studied their stabilities by a standard linearization
analysis, and only stable solutions are considered.

\begin{figure}
\includegraphics[width=3.25in]{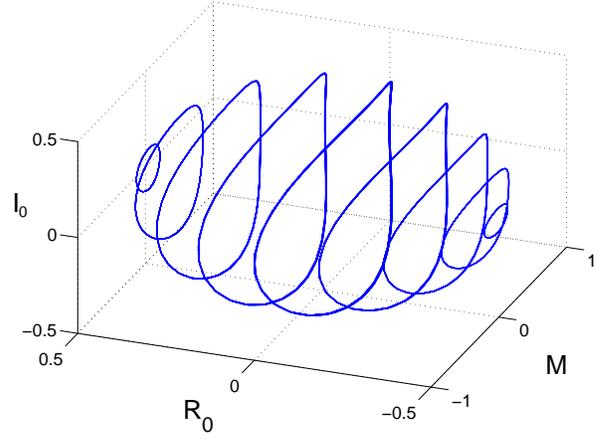}
\caption{The Josephson phase diagram for $J=0.02$ and $\lambda_A=-0.01$.
The trajectories of Eq. (\ref{jjd}) stay on closed paths. $\lambda_A<0$
($>0$) refers to a ferromagnetic (antiferromagnetic) spin-1 condensate
as for $^{87}$Rb ($^{23}$Na) atoms and $|\lambda_A|\ll|\lambda_S|$.}
\label{fig:fig1}
\end{figure}

\begin{figure}
\includegraphics[width=3.125in]{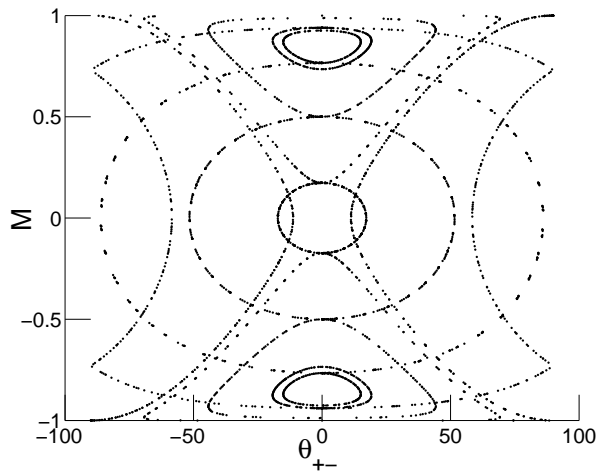}
\caption{The phase portrait for $J=0.0051$ and $\lambda_A= -0.01$.}
\label{fig:fig2}
\end{figure}

We now try to formulate the evolution of the most
easily measurable quantity: magnetization in each of the symmetric
double well. Since the total magnetization of our system is
a conserved quantity, we therefore only need to consider the
magnetization in the one of the wells, e.g. the left well.
We further consider a special
initial state $\eta_+=\xi_-$, $\eta_0=\xi_0$, and $\eta_-=\xi_+$.
The density matrix of the system then becomes
$\rho=\vec{\xi}^*\otimes\vec{\xi}^T$ with
$\rho_{ij}=\xi_i^*\xi_j$ for $i,j=+,0,-.$
It may be noted that $\rho$ stands for single-particle density matrix, as the many-body
density matrix is beyond the scope of a mean field analysis.
The dynamical variables contributing to the evolution of magnetization $M$ are given by
\begin{eqnarray}
M&=&\rho_{++}-\rho_{--};\quad n_0=\rho_{00};\nonumber\\
R_\pm&=&\frac{1}{2}(\rho_{+0}\pm\rho_{0-}+c.c);\quad R_0=\frac{1}{2}(\rho_{+-}+\rho_{-+});\nonumber\\
I_\pm&=&\frac{1}{2i}(\rho_{+0}\pm\rho_{0-}-c.c);\quad
I_0=\frac{1}{2i}(\rho_{+-}-\rho_{-+}). \quad \quad
\end{eqnarray}
We note that $R_0^2+I_0^2={[(1-n_0)^2+M)][(1-n_0)^2-M]}/{4}$, which comes
directly from the definitions of $R_0$, $I_0$, $M$, and the normalization
condition for $\vec{f}$. Quite generally, there is a conserved quantity, $R_+$,
in this system. Further restricting the initial state to $\xi_0=0$ we
find a simple situation analogous to the standard Josephson
junction for a scalar condensate in a double well \cite{milburn,raghavan99}
\begin{eqnarray}
\dot{M}&=&4JI_0,\nonumber\\
\dot{R_0}&=&-2\lambda_A I_0M,\nonumber\\
\dot{I_0}&=&2\lambda_A R_0M-JM.
\label{jjd}
\end{eqnarray}
This set of equations can be interpreted as the familiar
two state optical Bloch
equations, albeit nonlinear ones, with $R_0$ and $I_0$ being the
analogous real and imaginary parts of the atomic dipole moment.
$M$ then acts as the population inversion.
The fixed point $I_0=M=0$ is a center, which is not an attractor.
When $J\leq|\lambda_A|$, there exists another fixed point $I_0=2\lambda_A R_0-J=0$,
which is unstable since one of its three characteristic
frequencies is positive. Since $M$ is physically bounded between $1$ and $-1$,
we anticipate there exists a limit cycle in this system. For
$\xi_+=1$, we find $R_0=(1-M^2){\lambda_A}/{4J}$. The phase diagrams for
the case of full Josephson oscillations is shown in Fig.
\ref{fig:fig1}.
Alternatively, we can construct
two dimensional phase portraits by introducing an auxiliary variable
$\theta_{+-}=\tan^{-1}{(I_0/R_0)}$
as illustrated in Fig. \ref{fig:fig2}.

\begin{figure}
\includegraphics[width=3.125in]{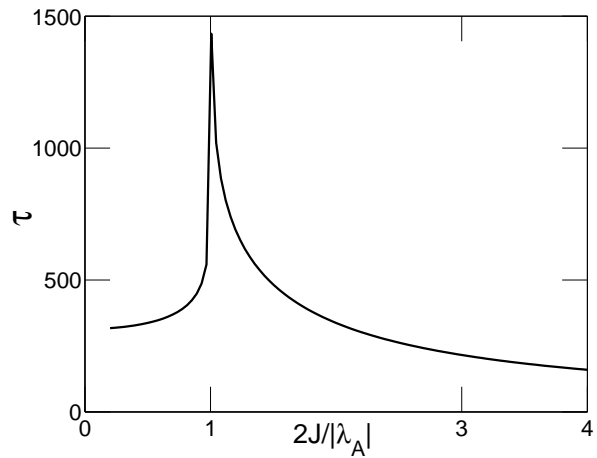}
\caption{Dependence of the period to the $2J/|\lambda_A|$.}
\label{fig:fig3}
\end{figure}

\begin{figure}
\includegraphics[width=3.25in]{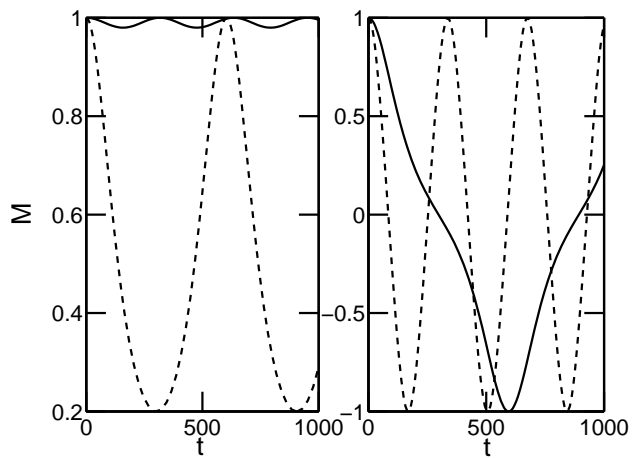}
\caption{Initial state is $\vec{\xi}(0)=(1,0,0)$ and
$\vec{\eta}(0)=(0,0,1)$. $\epsilon=1.0$, $\lambda_S=1.0$, and $\lambda_A=-0.01$. Left
figure shows self trapping when $2J\leq|\lambda_A|$, solid line for
$J=0.001$, dashed line for $J=0.0049$. Right figure shows
magnetization oscillation, solid line for $J=0.0051$, dashed
line for $J=0.01$.}\label{fig:fig4}
\end{figure}

Equation (\ref{jjd}) can be analytically solved
in terms of the Jacobi elliptic functions
\cite{raghavan99} with the oscillation period given by
\begin{eqnarray}
\tau&=&\left\{\begin{array}{c}
2K(\frac{2J}{|\lambda_A|})/{|\lambda_A|},\quad 2J<|\lambda_A|,\\
4K(\frac{|\lambda_A|}{2J})/{(2J)},\quad
2J>|\lambda_A|,\end{array}\right.
\label{eq:period}
\end{eqnarray}
where $K(.)$ is the complete elliptic integral of the first kind.
The behavior of the oscillation period with respect to $2J/|\lambda_A|$ is
shown in Fig. \ref{fig:fig3}. As we can see, for small values of
$2J/|\lambda_A|$ the gradually increasing period resembles that of
a nonlinear pendulum oscillation, which exhibits self-trapping.
The sel-trapping condition is therefore $2J<|\lambda_A|$.
$2J/|\lambda_A|=1$ is the critical value corresponding
to the homoclinic orbit of the equivalent pendulum being completely
in the top position. Beyond that the period decreases with increasing $2J/|\lambda_A|$,
as the equivalent pendulum assumes a librator rotation, i.e.
a full oscillation of the magnetization.
Figure \ref{fig:fig4} clearly illustrates these different behaviors for
the initial state $\vec{\xi}(0)=(1,0,0)$ and
$\vec{\eta}(0)=(0,0,1)$.

\begin{figure}
\includegraphics[width=3.125in]{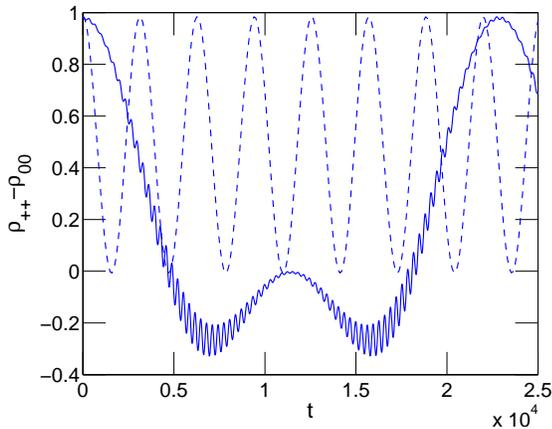}
\caption{Oscillation of the collective variable $\rho_{++}-\rho_{00}$
for an initial state $\vec{\xi}(0)=(0.9962,0.0872,0)$. Other parameters used are
$\epsilon=1.0$, $\lambda_S=1.0$, $\lambda_A=-0.01$ (solid line), and
$\lambda_A=0$ (dashed line for the case of no spin dependent atomic interaction)
, and $J = 0.001$.}
\label{fig:fig5}
\end{figure}

When the population in the Zeeman state $|M_F=0\rangle$ becomes non-zero,
even very small,
for instance with $\xi_+=\eta_-=0.9962$, $\xi_0=\eta_0=0.0872$, and $\xi_-=\eta_+=0$,
interaction caused self-trapping disappears,
while the full magnetization oscillation reappears in the previously
self-trapped regime. Yet, surprisingly, the presence of the
nonlinearity due to atomic self-interaction remains visible through
dynamic phenomena analogous to
self-trapping in the collective variables
of other combinations of the single particle density matrix elements.
In particular if we examine the collective variable
$\rho_{++}-\rho_{00}$,  we find that it does not always exhibit full oscillations.
 As an example, we demonstrate this
effect, unique to a spinor condensate, in Fig. \ref{fig:fig5}
for $\rho_{++}-\rho_{00}$.
With the given initial conditions full (complete) oscillations
between $1$ and $0$ occur when the spin dependent atomic self-interaction
is turned off, as shown by the dashed-line.
Inclusion of the self-interaction, on the other hand,
inhibits the full oscillation, leading to a similar
self-trapping effect as in the non-linear Josephson dynamics
of a scalar condensate in a double well.
As a general rule, we find oscillations now become incomplete
over any regular interval. The values of $\rho_{++}-\rho_{00}$
can even become negative as a result of including the self-interaction.
We can summarize the above observations based on the phase space nonlinear
dynamics. When $\lambda_A$ is non-zero and in the self trapping regime
when $2J<|\lambda_A|$, our model system
possesses one attractor, which attracts trajectories of
different collective variables depending on the initial conditions.

In the opposite regime to self-trapping when $2J>|\lambda_A|$,
the system possesses two distinct attractors, leading to a complete
oscillation of magnetization between $-1$ and $1$. We can adjust the
system parameters to shrink the distance between the two attractors
in the phase space, and generate beatings in the magnetization
oscillation with appropriately chosen initial conditions.
The mathematical reason is simple.
Due to the presence of the self-interaction in the
limit of $2J>|\lambda_A|$, the orbits of magnetization in the phase space
now become open and continuously circulate around two closely spaced focuses
(of the two attractors),
leading to beating patterns.

Finally, we provide some supporting arguments for the mean
field treatment of the tunnelling dynamics. The validity of the
mean field theory can be intuitively expected due to the large
numbers of atom in our model. At $N\sim 10^7$ atoms, our
system is beyond the microscopic regime, and becomes at least
mesoscopic if not completely macroscopic.
We note that the period of magnetization oscillation, given in Eq.
(\ref{eq:period}), has a simple dependence on the number of atoms
$\tau\sim 1/N$. Following the same analysis as in Ref.
\cite{zhang04}, we can estimate the effect of atom number
fluctuations
 by $|[\tau(N\pm\sqrt{N})-\tau(N)]/\tau({N})|\sim 1/\sqrt{N}$,
which is less than $0.003\%$ for the present system. We have simply
taken the standard deviation $\sigma(N)$ of the atom number
fluctuation to be $\sim\sqrt{N}$ consistent with the mean field
theory. We can also estimate the absolute phase diffusion times of
our system due to atom number fluctuations. The spin symmetric
interaction in a spin-1 condensate, gives rise to a diffusion time
of $\tau_c^{(S)}\sim 20$ (seconds) for the overall condensate
phase; while the spin asymmetric interaction, gives rise to a
diffusion time of $\tau_c^{(A)}\sim 5000$ (seconds) for the
relative phases between different condensate components. In the above
estimates, we have simply taken
$\tau_c^{(S)}\approx N/[\sigma(N)c_S\langle
n\rangle]$ as for a scalar condensate \cite{you} and
$\tau_c^{(A)}\approx N/[\sigma(N)c_A\langle n\rangle]$ for the
spin mixing dynamics \cite{law98}. We have also assumed an average
condensate density of $\langle n\rangle=1.7\times10^{13}
(\mathrm{cm})^{-3}$ for $N=10^7$ $^{87}$Rb atoms, which
corresponds to a typical Josephson oscillation period of $\sim 1$
(second). Thus we conclude that macroscopic magnetization
tunnelling in a spinor condensate as modelled here can be examined
faithfully with the mean field theory. We further note that in a
recent experiment with a smaller number of atoms $N\sim 1000$, the
nonlinear Josephson oscillations observed were in complete agreement with
the predictions of the mean field theory \cite{albiez04}.

In conclusion, we have studied the double well system of a spin-1
atomic condensate using the mean field theory and SMA.
We have characterized the various regimes of the
nonlinear dynamics for the
resulting macroscopic quantum tunnelling phenomena. In addition to features
commonly attached to the scalar Josephson dynamics of a single
component condensate in a double well, e.g. coherent atomic
population oscillation and macroscopic self-trapping
\cite{raghavan99,milburn}, we have found interesting effects
solely due to the spinor nature of the condensate such as the
macroscopic oscillation and self-trapping of condensate
magnetization without net changes of total atom numbers within each well.
The physics of these correlated tunnelling dynamics is
essentially the same as what was found before in a double well system
of a two component (or spin 1/2) condensate \cite{law}, but can become
significantly richer due to the additional freedom in atomic
internal states. For instance,
amplitude modulations in the magnetization oscillation become possible
in a spin-1 condensate.
We have illustrated the tunnelling induced
macroscopic magnetization oscillations with several examples for a
spin-1 condensate of ferromagnetic interactions. Our results
highlight the coherent Josephson dynamics of a spinor condensate
in a double well, and lay the ground work for future studies of
the quantum state of atoms in such a system.

We thank Dr. A. Smerzi for helpful comments and discussions.
O.E.M. acknowledges support from a T\"UBA/GEB\.{I}P award. The
work of L. Y. is supported by the NSF and NASA.


\begin{thebibliography}{}

\bibitem{leggett87} A. J. Leggett, S. Chakravarty, A. T. Dorsey,
M. P. A. Fisher, A. Garg, W. Zwerger, Rev. Mod. Phys. {\bf 59}, 1
(1987); A. J. Leggett, in {\it Chance and Matter}, eds. J.
Souletie, J. Vannimenus, and R. Stora (North-Holland, Amsterdam,
1987).

\bibitem{leggett85} A. J. Leggett and A. Garg, Phys. Rev. Lett.
{\bf 54}, 857 (1985).

\bibitem{garg93} A. Garg, Europhys. Lett. {\bf 22}, 205 (1993).

\bibitem{chudnovsky98} E. M. Chudnovsky and J. Tejada, {\it
Macroscopic Quantum Tunneling of the Magnetic Moment}, Cambridge
Studies in Magnetism, Vol. 4 (Cambridge University Press,
Cambridge, 1998).

\bibitem{gunther95} L. Gunther and B. Barbara, eds., {\it Quantum
Tunneling of Magnetization - QTM'94}, Vol. {\bf 301} of NATO Advanced
Study Institute, (Kluwer, Dordrecht, 1995).

\bibitem{villain94} J. Villain, F. Hartman-Boutron, R. Sessoli,
and A. Rettori, Europhys. Lett. {\bf 27}, 159 (1994).

\bibitem{lis80} T. Lis, Acta Crystallogr. B {\bf 36}, 2042
(1980).

\bibitem{wieghart84} K. Wieghart, K. Pohl, I. Jibril, and G.
Huttner, Angew. Chem. Int. Ed. Engl. {\bf 23}, 77 (1984).

\bibitem{awschalom92} D. D. Awschalom, J. F. Smyth, G. Grinstein,
D. P. DiVincenzo, and D. Loss, Phys. Rev. Lett. {\bf 68}, 3092
(1992); A. Garg, Phys. Rev. Lett. {\bf 70}, 1541 (1993).

\bibitem{kurn}D. M. Stemper-Kurn, M. R. Andrews, A. P. Chikkatur,
S. Inouye, H.-J. Miesner, J. Stenger, and W. Ketterle, Phys. Rev.
Lett. {\bf 80}, 2027 (1998).

\bibitem{mike} M. Barrett, J. Sauer, and M. S. Chapman,
Phys. Rev. Lett. {\bf 87}, 010404 (2001).

\bibitem{pu02}H. Pu, W. Zhang, P. Meystre, Phys. Rev. Lett. {\bf 89}, 090401 (2002).

\bibitem{davis}
A. Marchenkov, R. W. Simmonds, S. Backhaus, A. Loshak, J. C. Davis, and R. E. Packard,
Phys. Rev. Lett. {\bf 83}, 3860 (1999);
S. Backhaus, S. Pereverzev, R. W. Simmonds, A. Loshak, J. C. Davis,
and R. E. Packard,
Nature {\bf 392}, 687 (1998).

\bibitem{niu}Q. Niu, X.-G. Zhao, G. A. Georgakis, and M. G. Raizen
Phys. Rev. Lett. {\bf 76}, 4504 (1996);
S. R. Wilkinson, C. F. Bharucha, K. W. Madison, Q. Niu, and M. G. Raizen
{\it ibid}, 4512 (1996).

\bibitem{ens}M. Ben Dahan, E. Peik, J. Reichel, Y. Castin, and C. Salomon
Phys. Rev. Lett. {\bf 76}, 4508 (1996).

\bibitem{haycock00} D. L. Haycock, P. M. Alsing, I. H. Deutsch, J.
Grondalski, and P. S. Jessen, Phys. Rev. Lett. {\bf 85}, 3365
(2000).

\bibitem{mark}C. Orzel, A. K. Tuchman, M. L. Fenselau, M. Yasuda, and M. A. Kasevich,
Science {\bf 291}, 2386 (2001).

\bibitem{you}A. Imamoglu, M. Lewenstein, and L. You, Phys. Rev. Lett. {\bf 78}, 2511 (1997).

\bibitem{milburn}G. J. Milburn, J. Corney, E.M. Wright, and
D. F. Walls, Phys. Rev. A {\bf 55}, 4318 (1997).

\bibitem{sipe}R. W. Spekkens and J. E. Sipe,
Phys. Rev. A {\bf 59}, 3868 (1999).

\bibitem{ho}T. -L. Ho and C. V. Ciobanu,
J. Low. Temp. Phys. {\bf 135}, 257 (2004).

\bibitem{ashab01} S. Ashhab and C. Lobo, Phys. Rev. A {\bf 66}, 013609 (2002).

\bibitem{law}H. T. Ng, C. K. Law, and P. T. Leung,
Phys. Rev. A {\bf 68}, 013604 (2003).

\bibitem{leggett01} A. J. Leggett, Rev. Mod. Phys. {\bf 73}, 307
(2001).

\bibitem{ho98}T.-L. Ho, Phys. Rev. Lett. {\bf 81}, 742 (1998);
T.-L. Ho and S. K. Yip, Phys. Rev. Lett. {\bf 84},
4031 (2000).

\bibitem{law98} C. K. Law, H. Pu, and N. P. Bigelow, Phys. Rev. Lett.
{\bf 81}, 5257 (1998).

\bibitem{pu}H. Pu, C.K. Law, and N.P. Bigelow, Physica B {\bf 280},
27 (2000).

\bibitem{yi} S. Yi, O. E. M\"ustecapl{\i}o\u{g}lu, C. P. Sun,
 and L. You, Phys. Rev. A {\bf 66}, 011601 (2002).

\bibitem{raghavan99}A. Smerzi, S. Fantoni, S. Giovanazzi, and S. R. Shenoy
Phys. Rev. Lett. {\bf 79}, 4950-4953 (1997); S. Raghavan, A. Smerzi, S. Fantoni, and S. R.
Shenoy, Phys. Rev. A, {\bf 59}, 620 (1999); S. Giovanazzi, A.
Smerzi, and S. Fantoni, Phys. Rev. Lett. {\bf 84}, 4521 (2000).

\bibitem{zhang04} W. Zhang, D.L. Zhou, M.-S. Chang, M.S. Chapman, and L. You,
(to be published).
\bibitem{yi03} S. Yi,\"O. E. M\"ustecapl{\i}o\u{g}lu, and L. You,
Phys. Rev. Lett.{\bf 90}, 140404 (2003).
\bibitem{albiez04} M. Albiez, R. Gati, J. F\"olling, S. Hunsmann, M. Cristiani, and M.K. Oberthaler,
e-print cond-mat/0411757v2.


\end{thebibliography}
\end{document}